\documentclass[twocolumn,showpacs,preprintnumbers,amsmath,amssymb]{revtex4}
\usepackage{amssymb}
\usepackage{amsmath}
\usepackage{graphicx}
\usepackage{float}
\usepackage{dcolumn}
\usepackage{bm}
\usepackage{color}
\usepackage{exscale}
\usepackage{relsize}
\usepackage{extarrows}
\usepackage{xcolor}

\begin{document}

\title{Quantum Zeno effect in the multimode quantum Rabi model}
\author{Shu He$^{1,2,4}$}\email{heshu1987@foxmail.com}
\author{Chen Wang$^{3}$}\email{wangchenyifang@gmail.com}
\author{Xue-Zao Ren$^{2}$}
\author{Li-Wei Duan$^{4}$}
\author{Qing-Hu Chen$^{4}$}\email{qhchen@zju.edu.cn}
\address{
$^{1}$ Department of Physics and Electronic Engineering, Sichuan Normal University, Chengdu 610066, China \\
$^{2}$ School of Science, Southwest University of Science and Technology,
Mianyang 621010, China \\
$^{3}$ Department of Physics, Zhejiang Normal University, Jinhua, Zhejiang 321004, China \\
$^{4}$ Department of Physics, Zhejiang University, Hangzhou 310027, P. R. China
 }

 \date{\today }

\begin{abstract}
 We study the quantum Zeno effect (QZE) and quantum anti-Zeno effect (QAZE) of the multimode quantum Rabi model(MQRM). We derive an analytic expression for the decay rate of the survival probability where cavity modes are initially prepared as thermal equilibrium states. A crossover from  QZE to QAZE is observed due to the energy backflow induced by high frequency cavity modes. In addition, we apply a numerically exact method based on the thermofield dynamics(TFD) theory and the matrix product states(MPS) to study the effect of squeezing of the cavity modes on the QZE of the MQRM.  The influence of the squeezing angle, squeezing strength and temperature on the decay rate of the survival probability are discussed.
\end{abstract}

\maketitle

\section{Introduction}
Frequent measurements on a quantum system may cause its dynamical evolution slow down or accelerated. This phenomenon, known as the quantum Zeno effect(QZE) or quantum anti-Zeno effect (QAZE)\cite{misra1977zeno,facchi2008quantum,francica2010quantum,schafer2014experimental}, has been observed in experiments of trapped ion\cite{itano1990quantum}, cavity QED\cite{bernu2008freezing} and nuclear spin ensembles\cite{alvarez2010zeno}.
QZE has also been considered as a powerful strategy to implement the quantum control, including quantum communication\cite{pouyandeh2014measurement,bayat2015measurement}, quantum information protection\cite{barenco1997stabilization}, decoherence suppression\cite{beige2000quantum}, purification and cooling\cite{erez2008thermodynamical}.

Recently, the QZE has been observed in the circuit-QED system where a superconducting flux qubit is coupled to a transmission line\cite{kakuyanagi2015observation}. Compared to the traditional optical experimental domain, the time scales involved in the circuit-QED system are much larger and can be resolved with ordinary electronics. Moreover, the ultrastrong-coupling strength between the the qubit and the resonator has already been realized in circuit-QED systems\cite{forn2017ultrastrong}, making it possible to observe QZE and QAZE in the strong coupling regime.  As one of consequences in this regime, additional modes of the electromagnetic resonator become increasingly relevant\cite{bosman2017multi}. To capture such multiply modes effect, the widely used quantum Rabi model\cite{rabi1936process}, describing an qubit interacting with a single electromagnetic mode, has to be generalized to its multimode version, known as multimode quantum Rabi model(MQRM)\cite{sundaresan2015beyond,gely2017convergence}. Recent studies have shown that such higher-lying electromagnetic modes of the resonator has a profound impact on various quantum optical phenomenons in the strong coupling regime\cite{sanchez2018resolution,de2014light,garcia2015light,sundaresan2015beyond,gely2017convergence}.
Thus, it is an interesting question that what  role  do multiple modes play in the quantum Zeno dynamics of the MQRM.
In addition, previous studies have shown that the squeezing of the resonator mode has a significant influence in the quantum Zeno dynamics of a qubit in the quantum Rabi model\cite{lizuain2010zeno}. It is natural to raise a question that how does the squeezing of multiple resonator modes affect the QZE and QAZE of the MQRM.

%

To address the above problems, we exploit a numerical exact method based on the matrix product states and time-dependent variational principle\cite{haegeman2016unifying} to study the QZE and QAZE of the MQRM. By restricting the evolution in the single excitation subspace under the framework of TFD, we derive an analytical expression for the decay rate of survival probability when multiply modes are initially prepared as a thermal equilibrium states. We observe a crossover between the QZE to the QAZE under repeated projective measurements. By numerically calculating the energy transport between the TLS and multiply modes, we show that this crossover is attributed to the energy back flow from the high frequency modes to the qubit.
Moreover, we generalize the initial state of multiple modes to a squeezed thermal state and  study effects of squeezing phase angle and amplitude to the QZE of the MQRM. We find that the decay of survival probability is accelerated by non-vanishing squeezing strengthes. Squeezing angles also significantly affect the decay rate. Particularly, high frequency modes in the MQRM cause a positive shift on the critical squeezing angles where the decay rate reaches its extremum.

This paper is organized as follows. In Sec \uppercase\expandafter{\romannumeral2}, we briefly introduce the MQRM and the numerical exact method we used to  obtain  the evolution of the MQRM at finite temperature. In Sec \uppercase\expandafter{\romannumeral3}, we study QZE the MQRM at finite temperature. We discuss the crossover of the survival probability decay rate from QZE to QAZE and its relation to the energy transport between the qubit and multiple cavity modes.
Then, we extend the initial state of multiple cavity modes to squeezed thermal states and discuss the effect of the squeezing on the QZE in Sec \uppercase\expandafter{\romannumeral4}. We close this paper with a short summary in Sec \uppercase\expandafter{\romannumeral5}.

\section{Model and Methodology}

\subsection{The multimode quantum Rabi model}
In this paper, we focus on the MQRM which describes a qubit or a two-level system(TLS) interacting with multiple quantized photonic modes(e.g., cavity or resonator). The Hamiltonian of MQRM can be written as\cite{gely2017convergence,malekakhlagh2017cutoff}($\hbar =1$):
\begin{align}
 &H   = H_0 + H_B +H_{\text{int}} \notag\\
 &H_0 = \frac{\Delta}{2}\sigma_z, \quad  H_B = \sum_{m=0}^{M-1} \omega_m a_m^\dag a_m   \notag \\
 &H_\text{int} = \sum_{m=0}^{M-1}  g_m\sigma_x(a_m^\dag +a_m)\label{HamT}
\end{align}
where  $\sigma_x,\sigma_z$ are standard pauli operators and $a_m^\dag $,$a_m$ are the creation and annihilation boson operators for the $m$th mode with the frequency $\omega_m = (m+1)\omega_0$ and coupling strength $g_m =\sqrt{m+1}g$ to the TLS. $M$ is the total number of modes and depends on the specific physical realization. If not mentioned, we set $M=15$ throughout this study which is already feasible in the current circuit-QED experiment\cite{sundaresan2015beyond}. Since we are interested in the resonant situation, $\Delta = \omega_0$ is assume in this study.

To study the QZE of the MQRM, we focus on the survival probability of the TLS under successive ideal projective measurements with operator $\sigma_z$. The initial state is assumed as the product state: the TLS is at the exited state $|\uparrow\rangle $  and cavity modes are squeezed thermal states:
\begin{align}
 \rho(0) = |\uparrow\rangle\langle \uparrow|\otimes \rho_{\text{B}}, \quad \rho_{\text{B}} =  S(\xi)\left(Z(\beta)^{-1}e^{-\beta H_B} \right)S^\dag(\xi)  \label{IniOrigSq}
\end{align}
here $Z(\beta)=\text{Tr} e^{-\beta H_B}$ is the partition function. $\beta =1/(k_BT)$ where $T$ is the temperature and $k_B$ is the Boltzmann constant. $S(\xi) $ is the squeezing operator:
\begin{align}
 S(\xi) = \prod_{m}S_m(\xi_m) = \prod_m\exp\left(-\frac{1}{2}\xi^*_m a_m^2 +\frac{1}{2}\xi_m(a_m^\dag)^2 \right) \label{SqOp}
\end{align}
where $\xi_m= r_m{e^{i\phi_m}}$ is the squeezing parameter  for the $m$th mode. For simplicity, we assume a uniform squeezing for all $M$ modes, i.e. $\xi_m = \xi = re^{i\phi}$.

\subsection{MPS-based numerical method}

In the following, we briefly introduce the numerically exact method based on TFD and MPS to simulate the dynamics. We first remove the  squeezing operator in the initial state (\ref{IniOrigSq}) by applying the unitary transform $S(\xi)$ to the Hamiltonian ($\ref{HamT}$):
\begin{align}
  &H' = S^\dag(\xi) HS(\xi) = H_0 + H_B' + H_\text{int}' \notag\\
  & H_B' = \sum_{m=0}^{M-1}  \omega_m \left(Aa_m^\dag a_m + Ba_m^{\dag 2} + B^*a_m^{ 2}\right) \notag\\
  & H_\text{int}' = \sum_{m=0}^{M-1} g_m\left(Ka_m^\dag  + K^* a_m \right)\sigma_x \label{HamT2}
\end{align}
where $A = \cosh^2 r +\sinh^2 r, B = e^{i\phi }\cosh r\sinh r, K = \cosh r +e^{i\phi}\sinh r $.

According to the TFD method\cite{suzuki1991density,gelin2017thermal}, the evolution of this transformed Hamiltonian (\ref{HamT2}) from the initial state of thermalized cavity modes is equivalent to a Schr\"{o}dinger equation of a modified Hamiltonian $\hat{H}$ defined in an enlarged Hilbert space(see Appendix):
\begin{align}
  &\hat{H} = H_0 + \hat{H}_B + \hat{H}_{\text{int}} \notag\\
  &\hat{H}_B = \sum_{m=0}^{M-1} \omega_m\big[A \left(a^\dag_m a_m -b_m^\dag b_m\right) +\left( Ba_m^{\dag 2}-B b_m^2 +\text{H.c}\right)\big]  \notag\\
  &\hat{H}_{\text{int}} = \sum_{m=0}^{M-1} g_m(Ka_m^\dag \cosh\theta_m +Kb\sinh\theta_m)\sigma_x + \text{H.c} \label{Ham3}
\end{align}
The {the vacuum initial state} is
\begin{align}
 |\hat{\psi}(0)\rangle = |\uparrow\rangle \otimes \prod_{m=0}^{M-1}|0\rangle_{a_m}|0\rangle_{b_m} \label{Inistate2}
\end{align}
where $b_m^\dag,b_m$ are boson operators of the fictitious modes and $\theta_m = \text{arctanh}(e^{-\beta\omega_m/2}) $ . The evolution of the expectation value of an arbitrary operator $O$ that affects in original Hilbert space  can be straightforwardly calculated :
\begin{align}
 \langle O(t)\rangle =  \langle \hat{\psi}(t)|O |\hat{\psi}(t)\rangle
\end{align}

Finally, the TFD Schr\"{o}dinger equation is simulated by MPS-based numerically exact method \cite{schollwock2011density,haegeman2016unifying,wall2016simulating} which has been widely used and proved to be highly efficient in solving quantum many body dynamics\cite{schroder2016simulating,kloss2018time}.

\subsection{Survival probability and effective decay rate}

Now we turn to the QZE and QAZE of the MQRM.  The QZE \cite{misra1977zeno} can be described by the survival probability $P_\text{sur}(n\tau)$ which is defined as the probability of finding the initial state after $n$ successive measurements with equal time interval $\tau$. The measurement considered in this paper is assumed to be an ideal projecting measurements of operator $\sigma_z$ followed by a post selection regarding to the positive measurement result.The survival probability can be written as:
\begin{align}
 P_{\text{sur}}(n\tau) = \text{Tr}\left[ P_{}e^{-i\hat{H}\tau}\rho(0)e^{i\hat{H}\tau}P_{} \right]^n = P^n_{\text{sur}}(\tau) \label{PsurWeak}
\end{align}
where $P_{} = |\uparrow\rangle\langle \uparrow|$ is measurement projecting operator. In the short interval time limit $\tau \rightarrow 0$,  one can further write $P_{\text{sur}}(t=n\tau)$ in an exponentially decay form\cite{facchi2001quantum,maniscalco2006zeno}:
\begin{align}
 P_\text{sur}(n\tau) = \exp(-\gamma(\tau)t)
\end{align}
where $\gamma(\tau)$ is  an effective decay rate:
\begin{align}
 \gamma(\tau) = -\frac{1}{\tau}\ln\left[P_\text{sur}(\tau)\right]\label{GammaLog}
\end{align}
Eq.(\ref{GammaLog}) is valid when the measurement interval $\tau$ is relatively short where the measurement disturbance to the environment(here is cavity modes) can be neglected thus the state after each projecting measurement collapse to the identical state as Eq.~(\ref{Inistate2})\cite{he2017zeno}.  Therefore, we restrict $g\tau <1$ in the following discussion.

The effective decay rate $\gamma(\tau)$ is a crucial quantity to characterize the QZE and the QAZE\cite{zheng2008quantum,cao2010dynamics}:
${\partial \gamma(\tau)}/{\partial \tau}>0$ means that the system is more severely slowed-down by faster repeated measurements, indicating the occurrence of QZE; on the contrary, ${\partial \gamma(\tau)}/{\partial \tau}<0$ can be regarded as the characteristic of QAZE since the decay is accelerated by frequent measurements. Compared to the original criterion to classify QZE by using $\gamma(\tau)/\gamma_0 <0 $ where $\gamma_0= \gamma(\tau\rightarrow \infty)$ is the natural decay rate\cite{zheng2008quantum}, new definitions through ${\partial \gamma}/{\partial \tau}$ retains the core physical picture of QZE and QAZE without calculating $\gamma_0$ that may not exist in some models\cite{chaudhry2014zeno,zhang2015zeno,wu2017quantum,he2017zeno}. Throughout this paper, we use this new criterion to classify QZE and QAZE, the potential crossover point is denoted as $\tau_c$, i.e. ${\partial \gamma(\tau)}/{\partial \tau}\big|_{\tau_c}=0$ .

\section{Results and discussion}

\subsection{Thermal equilibrium initial state}
We first consider the QZE when cavity modes are prepared as the thermal equilibrium state.
The survival  probability of Hamiltonian ($\ref{Ham3}$) with the product initial state ($ \ref{Inistate2}$) can be written as\cite{lizuain2010zeno}:
\begin{align}
 P_\text{sur}(\tau)& = \text{Tr}\big[\rho(\tau)|e\rangle\langle e|\big] = \text{Tr}\big[e^{-i\hat{H}\tau}\rho(0)e^{i\hat{H}\tau}|e\rangle\langle e|\big] \notag\\
 & = \text{Tr}\left[ \sum_{n=0}^\infty \frac{(-i\tau)^2}{n!}[\hat{H},\rho(0)]_n |e\rangle \langle e|\right]\label{PSum}
\end{align}
where $\rho_0  = |\hat{\psi}(0) \rangle \langle \hat{\psi}(0)|$ and $[\hat{H},\rho(0)]_n = [\hat{H},[\hat{H},\rho(0)]_{n-1}]$. A commonly used approximation to obtain the decay rate of the survival probability is to keep terms in (\ref{PSum}) $\text{}$ up to $\tau^2$ , this leads to the decay of $P_\text{sur}(\tau)$ in a quadratical form:
\begin{align}
 P_e(\tau) \approx 1-\left(\frac{\tau}{\tau_z}\right)^2
\end{align}
where $\tau_z $ is known as "Zeno time"\cite{schulman1994characteristic}. Thus for $N$ repeated projective measurement with equal interval  $\tau$, the survival probability can be approximated as an exponential decay:
\begin{align}
P(t=N\tau) = \bigg[1-\left(\frac{\tau}{\tau_z}\right)^2\bigg]^{N} \approx e^{-\gamma(\tau) t}
\end{align}
with the effective decay rate $\gamma_{\text{2nd}}(\tau)$:
\begin{align}
  \gamma_{\text{2nd}}(\tau) = \tau\sum_m g_m^2\left(\cosh^2(\theta_m) +\sinh^2(\theta_m)\right) \label{gamma2nd}
\end{align}

\begin{figure}[tbp]
\includegraphics[scale=0.64]{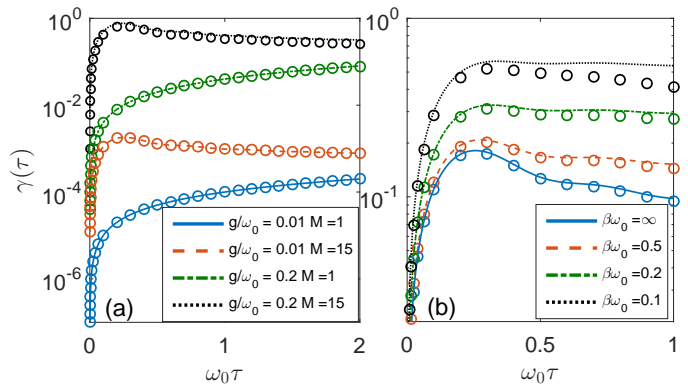}
\caption{(Color online) Effective decay rate of the survival probability $\gamma(\tau)$ for thermalized initial states. Results with lines are obtained from Eq.(\ref{AnaEq}) while circles are results from  numerical exact MPS-TDVP method.  (a) Comparison of decay rates between single mode and multiple modes at zero temperature.  (b)effective decay rate at different temperature for $g/\omega_0 =0.1$.}
\label{Fig1}
\end{figure}
Since Eq.(\ref{gamma2nd}) is independent of $\Delta$ and $\omega_m$ , its validity is limited to the case where the measurement interval $\tau$ is much shorter than the typical time scale of \emph{each mode} , that is $\tau<1/\omega_m$.
When high frequency modes are included, it severely precludes the availability of $\gamma_{\text{2nd}}(\tau)$ compared to the Rabi model where only single resonant mode is considered.

Instead, we employ an alternative method based on the thermofield dynamics(TFD)\cite{umezawa1982thermo,suzuki1991density}. This method transforms the evolution of a mixed thermalized initial state into another evolution with a pure initial state in an enlarged Hilbert space whose total number of degrees of freedom is double compared to the original one. Recently, this method has received much attention and has been widely used to study finite temperature dynamics of quantum electron-vibrational systems\cite{borrelli2016quantum,borrelli2017simulation,chen2017finite} and open quantum systems\cite{wang2017finite,he2018zeno}. Details of TFD can be found in Ref\cite{borrelli2016quantum,gelin2017thermal}.

By restricting the evolution of the TFD Schr\"{o}dinger equation to the single expiation subspace, we can obtain an analytic expression for the decay rate of the survival probability(see Appendix):
\begin{align}
  \gamma_{\text{th}}(\tau) =& \tau \sum_{m=0}^M\cosh^2 \theta_m g_m^2\text{sinc}^2\left[\frac{\tau(\omega_k-\Delta)}{2}\right] \notag\\
  +&\sinh^2\theta_m g_m^2\text{sinc}^2\left[\frac{\tau(\omega_k+\Delta)}{2}\right]\label{AnaEq}
\end{align}
where $\text{sinc(x)}\equiv \sin(x)/x$. Different from the $\gamma_{\text{2nd}}$ in Eq(\ref{gamma2nd}), $\gamma_{\text{th}}$ shows apparent dependence of frequencies $\omega_k$ and $\Delta$  and has a leading scaling of $g_k^2/(\Delta -\omega_k)^2$ which is similar to the  decay rate of the spontaneous emission in the  multimode Percell effect due to non-resonant modes\cite{houck2008controlling}.

In Figure(\ref{Fig1}), the analytical result $\gamma_{\text{th}}(\tau)$ agree well with numerical exact ones in a large range of parameters including strong coupling and high temperature regimes. Specifically, for zero temperature case( Figure(\ref{Fig1}.a)) the decay rate $\gamma(\tau)$ for a single mode Rabi model shows a monotonic increasing with the increase of measurement interval $\tau$ for both weak($g/\omega_0 = 0.01$) and strong($g/\omega_0 = 0.2$) coupling strengthes, clearly demonstrates a pure Zeno effect. However, by including the high-lying photonic modes($M=15$), the decay rate of survival probability presents a nonmonotonic behavior, which increases with $\tau$ before $\omega_0\tau<0.2$ and then decreases, indicating a crossover from QZE to QAZE.
Similar transitions can be observed for finite temperature situations, as shown in (Figure(\ref{Fig1}.b)). Thus we conclude that these high-frequency cavity modes in the MQRM may induce the QAZE. In the following, we show that such QAZE  is attributed to the energy backflow from  high-frequency modes to the TLS.

\begin{figure}[tbp]
\includegraphics[scale=0.66]{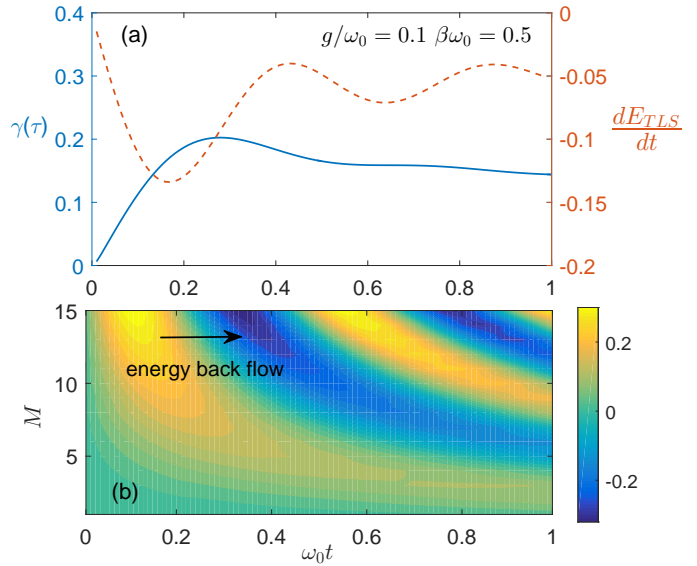}
\caption{(Color online) Energy transport rate between the TLS and multiple modes. (a) Decay rate $\gamma(\tau)$ and total energy transport rate of the TLS at $g\omega_0 =0.1,\beta\omega_0 =0.5$.  (b) Evolution of mode's energy operator $\langle E_m(\tau)\rangle$ with same parameters as (a). \label{Fig2}}
\end{figure}

According to Eq.(\ref{GammaLog}), the decay rate $\gamma(\tau)$ can be related to the population of the TLS in the excited state $|\uparrow\rangle$. Since the energy of the TLS is $E_{\text{TLS}}(t) = \Delta\langle \sigma_z(t)\rangle/2$ which is proportional to the energy of the TLS:
\begin{align}
 P_\text{sur}(\tau) = (\langle\sigma_z(\tau)\rangle+1)/2 =E_{\text{TLS}}(\tau)/\Delta+1/2
\end{align}
thus we may understand this crossover from the view of energy transport between multiple mods and the TLS. Since the QAZE is classified by $d\gamma(\tau)d\tau <0$, this requires:
\begin{align}
  \frac{d\gamma(\tau)}{d\tau} <0 \Leftrightarrow -\frac{1}{\tau}\left(\frac{d\ln(P_{\text{sur}})}{d\tau} -\frac{\ln(P_{\text{sur}})}{\tau}\right)< 0 \label{QAZE}
\end{align}
By considering the assumption that the measurement interval is short ($g\tau<1$), $E_{\text{TLS}}(\tau)$ can be approximately expressed as:
\begin{align}
 E_\text{TLS}(\tau)/\Delta = 1/2- a_1 (g\tau) -a_2 (g\tau)^2 +O(g\tau)^2 \label{ETLS}
\end{align}
where $a_1>0$ since the TLS is initially prepared in exited state and $E_{\text{TLS}}$ is maximized at $\tau=0$. Inserting Eq.(\ref{ETLS}) into Eq.(\ref{QAZE}), we obtain:
\begin{align}
 \frac{d\gamma(\tau)}{d \tau} \approx \frac{d\gamma}{d\tau}\bigg|_{\tau=0} +\frac{4}{3}\tau a_1g\left(\frac{d\gamma}{d\tau}\bigg|_{\tau=0} +\frac{1}{2}a_2g^2 \right) +O(\tau)<0
\end{align}
For arbitrary system, ${d\gamma}/{d\tau}|_{\tau=0}> 0 $ which implicates the pure Zeno effet when $\tau\rightarrow 0$, the occurrence of the QAZE  requires $a_2<0 $, i.e.  ${d^2E_{\text{TLS}}}/d^2\tau >0$. As depicted in Figure(\ref{Fig2}).a, the crossover point $\tau_{c}$  is close to
the pole of the energy transport rate as $d^2 E_{\text{TLS}}/d^2\tau =0$. As shown in $(\ref{Fig2})$, by further calculating the  evolution of the energy operator for each mode $E_m(\tau) = \omega_m\langle \psi(\tau)|a_m^\dag a_m|\psi(\tau)\rangle$, we observe an energy backflow from the higher  frequency parts of cavity modes to the TLS during the crossover point $\tau_c$. \emph{It is this energy  backflow induced by high frequency modes that leads to  $d^2 E_{\text{TLS}}/d^2\tau >0$ and makes it possible to observe QAZE.} On the contrary, the low frequency part of modes always absorbs energy from the TLS during $g\tau<1$, leading to the pure QAZE in the single mode Rabi model as shown in Figure(\ref{Fig1}).a.

\subsection{Squeezed thermal initial states}
In this section, we generalize the initial state of the cavity modes to the \emph{squeezed thermal states}.

From the view of the Hamiltonian (\ref{Ham3}), effects of the squeezing come from two aspects: 1).The renormalization of mode frequencies and effective coupling strengthes by factors $A$ and $K$ respectively.  2).The nonharmonic terms in $\hat{H}_B$ with the factor $B$.

\begin{figure}[tbp]
\includegraphics[scale=0.62]{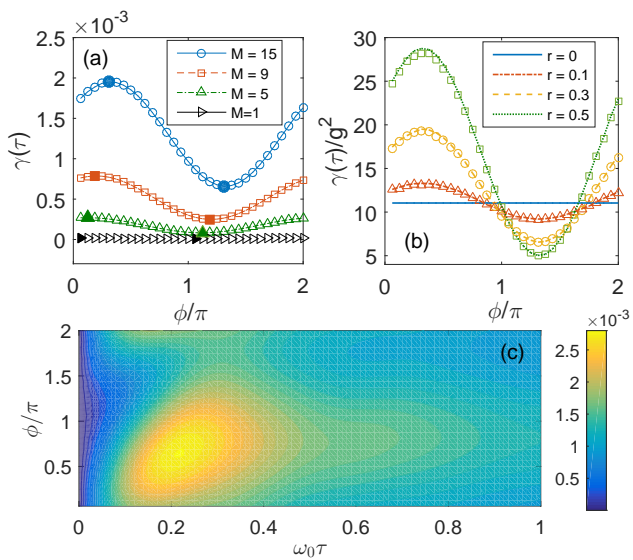}
\caption{(Color online) Relation between squeezing angles and the decay rates of survival probability $\gamma(\tau)$. (a). Decay rates of different squeezing angles for various numbers of cavity modes at zero temperature, the maximal and minimal critical angles are marked by solid symbols, other parameters are $g/\omega_0 = 0.01, \omega_0\tau =0.1, r=0.3$. (b). Decay rates as a function of squeezing angles at zero temperature for different squeezing strength $r$, data denoted by lines are obtained from $g/\omega_0 =0.01$ while hollows are obtained from $g/\omega_0 =0.1$. Other parameters are $\omega_0\tau =0.1 $. (c) Decay rates for different measurement interval $\tau$  and squeezing angle $\phi$, other parameters are $g/\omega_0 =0.01, r =0.3$. }\label{Fig3}
\end{figure}

The effect of 1) can be included in the analytic expression ($\ref{AnaEq}$) by replacing $g_m^2 \rightarrow |K|^2g_m^2$ and $\omega_m \rightarrow A\omega_m $. By increasing $r$, the effective coupling strength is increased, leading to a larger decay rate.  Moreover, due to the factor $|K|^2$, $\gamma(\tau)$  depends on the squeezing angle $\phi$. Since $|K|^2 = \cosh(2r) + \cos\phi \sinh(2r)$, the critical angle where $\gamma(\tau)$ takes the extreme value may be  $\phi_{\text{max}} = 0$ and $\phi_{\text{min}} = \pi$ respectively. Same critical angles have been observed on the single-mode rabi model with initial squeezed field\cite{lizuain2010zeno} and the TLS in a squeezed bath under RWA\cite{mundarain2006zeno}. However, this is not the case in the current MQRM where effects of both multimodes  and non-rotating terms are taken into consideration.
Specifically, by increasing the number of modes from $M =1$ to $M = 15$,  critical angles $\theta_\text{min}$ and $\theta_\text{max}$  gradually increase with a fixed angle difference $\theta_\text{min} - \theta_\text{max} = \pi$ as shown in Fig(\ref{Fig3}.a). Interestingly, as depicted Fig(\ref{Fig3}.b), such critical angles are independent of squeezing strength $r$ and coupling strength $g$. This independence agrees with results of the TLS in squeezed bath\cite{mundarain2006zeno}. Surprisingly, the scaling of  $\gamma(\tau)\sim g^2$ for thermal equilibrium initial states shown in Eq(\ref{AnaEq}) may still holds for squeezed thermal bath by noting that  $\gamma(\tau)/g^2$ as a function of $\phi$  for different coupling strengthes $g$ in Fig(\ref{Fig3}.b) collapses to the same curve.

The measurement interval $\tau$  also affects such critical angles, as shown in Fig(\ref{Fig3}.c). Note when $\tau\rightarrow 0$, the shift of critical angles approaches zero. The crossover from QZE to QAZE  can still be observed for the squeezed thermal initial state. In addition, a specific squeezing angle can either highlights or downplay this transition: as shown in Fig(\ref{Fig3}.c), during transition point $\tau_c$, $\gamma(\tau)$ are significantly increased for $\phi = \phi_\text{max}$ and suppressed for $\phi = \phi_\text{min}$, while with the increase of $\tau$, the difference between $\gamma(\tau)$ at two critical angles is narrowed.

\begin{figure}[tbp]
\includegraphics[scale=0.66]{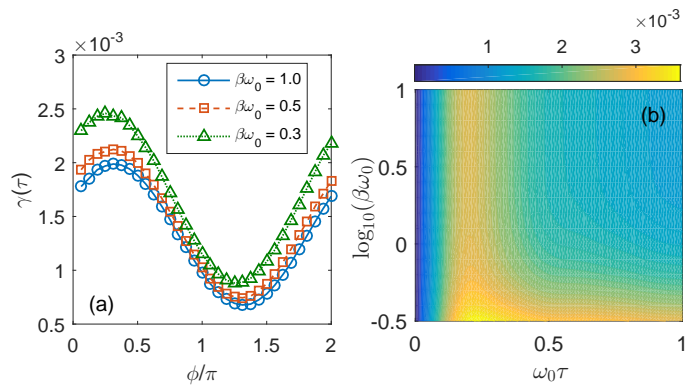}
\caption{(Color online)  Finte temperature effect on the decay rate $\gamma(\tau)$. (a) $\gamma(\tau)$ as a function of squeezing angle $\phi$ at different temperatures with $\omega_0\tau =0.1$. (b)  The decay rate $\gamma(\tau)$ for various measurement interval $\tau$ and temperature, $\phi = \pi/2$. For both subplots above, $g/\omega_0 =0.01$ . For both subplots, $g/\omega_0 = 0.01$, $r=0.3$. }
\label{Fig4}
\end{figure}

Finally, the effect of temperature on the decay rate for the squeezed thermal initial state is described by $\theta_m$ in $\hat{H}_{\text{int}}$ where higher temperature leads to larger effective coupling strength and consequently larger decay rate, as shown Fig(\ref{Fig4}.b). Moreover, increasing temperature also causes reduction of the shift of critical angles. This is not a surprising result. As explained above, the shift of critical angles are related to the $Ba_m^{\dag,2} + B^*a_m^{2}$ in $H_B$. However, with the increase of temperature, $\hat{H_\text{int}}$ becomes dominant compared to the  nonharmonic term, thus can be neglected in the high-temperature limit and the relation between $\gamma(\tau)$ and $\phi$ are only determined  through $|K|^2$.

\section{Conclusion}

In summary, we study the QZE and QAZE of the MQRM where cavity modes are prepared in thermal equilibrium state.  We derive an analytic expression of the decay rate of the survival probability  which shows
a transition from QZE to QAZE. By calculating the evolution of energy flow between the TLS and cavity modes, we show that such QZE-QAZE transition is related to the energy backflow from high frequency modes to the TLS. Furthermore, we generalize the initial state of cavity modes to squeezed thermal states and study effects of squeezing angle and strength by applying a numerically exact method. We find that the decay of survival probability is accelerated by non-vanishing squeezing strengthes. Moreover, squeezing angles also significantly affect the decay rate:  compared to $\theta_\text{max} =0$ and $\theta_\text{min} = \pi $ in which decay rates take maximal and minimal values in the single-mode Rabi model, high frequency modes in the MQRM cause a positive shift on these critical squeezing angles. Finally, with the increase of the temperature, the decay of the survival probability is accelerated and the shift of critical squeezing angles are reduced.


\section{Acknowledgement}
Shu He is supported by the National Natural Science Foundation of China under Grant No. 11804240.  Chen Wang is supported by the National Natural Science Foundation of China under Grant Nos. 11704093 and 11547124. Liwei Duan and Qing-Hu Chen are supported by the National Natural Science Foundation of China under Grant Nos.11674285 and 11834005.

\appendix

\section{A Brief introduction to TFD}
We briefly introduce the theory of thermofield dynamics in this Appendix.

The time evolution of an arbitrary Hamiltonian $H$ at finite temperature can be described  by the Liouville-von Neumann equation($\hbar=1$):
\begin{align}
 \partial_t \rho(t) = -i[H,\rho(t)]\label{rhoLV}
\end{align}
where the initial state is assumed to be at thermal equilibrium:
\begin{align}
 \rho(0) = \frac{1}{Z(\beta)}\exp\left(-\beta H\right)\label{IniRho}
\end{align}
here $Z(\beta)=\text{Tr}e^{-\beta H}$ is the partition function and $\beta = \left(k_BT\right)^{-1}$. $k_B$ is the Boltzmann constant and $T$ is the temperature.

According to Ref\cite{suzuki1986thermo}, the time evolution of such density matrix (\ref{rhoLV}) can be equivalently described by a modified Hamiltonian $\hat{H} = H-\tilde{H}$ and corresponding Schr\"{o}dinger equation defined in an enlarged Hilbert space($\hbar =  1$):
\begin{align}
 i\frac{\partial}{\partial t}|\psi(t)\rangle = \hat{H} |\psi(t)\rangle
\end{align}
$\tilde{H}$  represents the fictitious Hamiltonian which can be derived from the original Hamiltonian $H$\cite{suzuki1991density}.

By defining the vector $|I\rangle$:
\begin{align}
 |I\rangle = \sum_k|k\rangle |\tilde{k}\rangle
\end{align}
where $|k\rangle$($|\tilde{k}\rangle$) are arbitrary basis vectors of the original(fictitious) space, the density matrix $\rho(t)$ and wave function $|\psi(t)\rangle$ have the following relation\cite{suzuki1985thermo}:
\begin{align}
 |\psi(t)\rangle = \rho(t)^{1/2}|I\rangle
\end{align}
Particularly, the initial state at thermal equilibrium is:
\begin{align}
 |\psi(0)\rangle =\rho(0)^{1/2}|I\rangle = Z(\beta)^{-1/2}\exp\left( -\frac{1}{2}\beta H\right)|I\rangle
\end{align}

The expectation value for an arbitrary operator $A$ defined in the original Hilbert space $\{|k\rangle\}$ can be calculated by:
\begin{align}
 \langle A(t)\rangle = \langle \psi(t)|A |\psi(t)\rangle = \text{Tr}\left\{ \rho(t) A\right\}
\end{align}
Thus the evaluation of $\langle A(t) \rangle$ by using the TFD wave function $|\psi(t)\rangle$ is equivalent to the that by using the density matrix $\rho(t)$.

Now let us turn to the Hamiltonian of the MQRM (\ref{HamT}). To describe its evolution from the initial state $(\ref{IniOrigSq})$ under the framework of the TFD, we first remove the  squeezing operator in the initial state by applying a unitary transform $S(\xi)$ defined in (\ref{SqOp}) to the Hamiltonian ($\ref{HamT}$):
\begin{align}
  &H' = S^\dag(\xi) HS(\xi) = H_0 + H_B' + H_\text{int}' \notag\\
  & H_B' = \sum_{m=0}^{M-1}  \omega_m \left(Aa_m^\dag a_m + Ba_m^{\dag 2} + B^*a_m^{ 2}\right) \notag\\
  & H_\text{int}' = \sum_{m=0}^{M-1} g_m\left(Ka_m^\dag  + K^* a_m \right)\sigma_x \label{HamT2Ap}
\end{align}
where $A = \cosh^2 r +\sinh^2 r, B = e^{i\phi }\cosh r\sinh r, K = \cosh r +e^{i\phi}\sinh r $ , this is the Hamiltonian (\ref{HamT2}) in the main text.

We can choose the fictitious Hamiltonian $\tilde{H}$ as :
\begin{align}
 \tilde{H} = -\sum_m \omega_m\left( A b^\dag_m b_m + B^* b^{\dag 2}_m +Bb_m^2 \right)
\end{align}
where $b_m^\dag $($b_m$) is the creation (annihilation) boson operator defined in the fictitious Hilbert space corresponding to the the mode $a_m$. This leads to the total Hamiltonian for the TFD schr\"{o}dinger equation:
\begin{align}
 \hat{H} =& \frac{\Delta}{2}\sigma_z + \sum_m \omega_m\left( A a^\dag_m a_m + Ba_m^{\dag 2} + B^* a_m^{2}\right) \notag\\
 & -\sum_m \omega_m\left(A b^\dag_m b_m +B^*b_m^{\dag 2} + Bb_m^{2} \right)\notag\\
& + \sum_{m=0}^{M-1} g_m\left(Ka_m^\dag  + K^* a_m \right)\sigma_x  \label{HTotal}
\end{align}

The corresponding initial TFD wave function can be written as:
\begin{align}
 |\psi(0)\rangle =|\uparrow\rangle\otimes|\textbf{0}(\beta)\rangle
\end{align}
where $|\textbf{0}(\beta)\rangle$ is often referred to as the thermal vacuum state:
\begin{align}
 |\textbf{0}(\beta)\rangle = Z(\beta)^{-1/2}e^{-\beta H_B/2}|I\rangle  \label{IniA1}
\end{align}

This thermal vacuum state can be further transformed to the ground state(vacuum state) of the modes $\{a_m\}$ and $\{b_m\}$ by applying the so-called Bogoliubov thermal transformation\cite{takahashi1996thermo}:
\begin{align}
|\textbf{0}(\beta)\rangle =  e^{-iG}|\textbf{0}\rangle
\end{align}
where $|\textbf{0}\rangle = \prod_{m=0}^{M-1}|0\rangle_{a_m} |0\rangle_{b_m}$ and the transformation operator $G$ is defined as:
\begin{align}
 G = -i \sum_m\theta_m(a_mb_m - a^\dag_m b^\dag_m)
\end{align}
where
\begin{align}
 \theta_m  = \text{arctanh}(e^{-\beta\omega_m/2})
\end{align}

Instead of  solving the TFD schr\"{o}dinger equation (\ref{HTotal}) with the initial state (\ref{IniA1}) ,  one can apply the inverse Bogoliubov thermal transformation to the Hamiltonian $\hat{H}$. By considering the  following new relations:
\begin{align}
   &e^{iG} a_me^{-iG}  = a_m\cosh(\theta_m) + b_m^\dag \sinh(\theta_m) \notag\\
   &e^{iG} b_me^{-iG}  = b_m\cosh(\theta_m) + a_m^\dag\sinh(\theta_m)
\end{align}

We have:
\begin{align}
 &|\hat{\psi}(t)\rangle = e^{iG}|\psi(t)\rangle,\quad i\frac{\partial}{\partial t}|\hat{\psi}(t)\rangle = \hat{H}_\theta |\hat{\psi}(t)\rangle
\end{align}
 $\hat{H}_{\theta} = e^{iG}\hat{H}e^{-iG}$  is the final Hamiltonian (\ref{Ham3}) in the main text:
\begin{align}
  &\hat{H}_\theta = H_0 + \hat{H}_B + \hat{H}_{\text{int}} \notag\\
  & H_0   = \frac{\Delta}{2}\sigma_z \notag\\
  &\hat{H}_B = \sum_{m=0}^{M-1} \omega_m\big[A \left(a^\dag_m a_m -b_m^\dag b_m\right) +\left( Ba_m^{\dag 2}-B b_m^2 +\text{H.c}\right)\big]  \notag\\
  &\hat{H}_{\text{int}} = \sum_{m=0}^{M-1} g_m(Ka_m^\dag \cosh\theta_m +Kb\sinh\theta_m)\sigma_x + \text{H.c} \label{HTTT}
\end{align}
and corresponding initial state (\ref{Inistate2}):
\begin{align}
 |\hat{\psi}(0)\rangle = |0\rangle\otimes |\textbf{0}\rangle = |\uparrow\rangle \otimes \prod_{m=0}^{M-1}|0\rangle_{a_m}|0\rangle_{b_m} \label{INITTT}
\end{align}

\section{Derivation of Eq.(\ref{AnaEq}) }

In this Appendix, we show the detailed derivation to the analytic expression of the decay rate $\gamma_{\text{th}}(\tau)$.

The evolution of the MQRM from a product initial state where resonator modes are at thermal equilibrium is govern by the  TFD Schr\"{o}dinger equation of the Hamiltonian ($\ref{HTTT}$) with $r=0$(without squeezing):
\begin{align}
 &\hat{H}_{\text{th}}  =\frac{\Delta}{2}\sigma_z + \sum_m\omega_m( a_m^\dag a_m - b_m^\dag b_m )\notag\\
 &+ g_m\bigg(\cosh\theta_m(a_m^\dag +a_m) +\sinh\theta_m(b_m^\dag +b_m)\bigg)\sigma_x \label{HamiSE}
\end{align}

To obtain an analytic expression, we  further apply an approximation by restricting the evolution in the single excitation(SE) Hilbert space. This can be achieved by writing the wave function at arbitrary time $t$ as:
\begin{align}
 &|\psi(t)\rangle = \chi(t)|\uparrow\rangle|\textbf{0}\rangle \notag\\
 &+\sum_m p_m(t)|\downarrow\rangle|1\rangle_{a_m} |0\rangle_{b_m} +\sum_m q_m(t)|\downarrow\rangle|0\rangle_{a_m}|1\rangle_{b_m} \label{WaveFuncApp}
\end{align}

Thus the Hamiltonian (\ref{HamiSE}) can be simplified as:
\begin{align}
 &\hat{H}_{\text{th,SE}}  =\frac{\Delta}{2}\sigma_z + \sum_m\omega_m( a_m^\dag a_m - b_m^\dag b_m )\notag\\
 &+ g_m\bigg(\cosh\theta_m(\sigma_-a_m^\dag +\sigma_+a_m) +\sinh\theta_m(\sigma_-b_m^\dag +\sigma_+b_m) \bigg) \label{HamiFTRWA}
\end{align}
where $\sigma_\pm = \frac{1}{2}\left(\sigma_x\pm i\sigma_y\right)$.

Inserting the wave function (\ref{WaveFuncApp}) into the TFD Schr\"{o}dinger equation, we have:
\begin{align}
  &i\frac{d\chi(t)}{dt} = \frac{\Delta}{2}\chi(t) + \sum_m g_m\cosh\theta_m p_m(t) + g_m\sinh\theta_mq_m(t)\notag\\
 &i\frac{dp_m(t)}{dt}=\sum_m (\omega_m - \frac{\Delta}{2})p_m(t) + g_m\cosh\theta_m\chi(t)\notag\\
 &i\frac{dq_m(t)}{dt}=\sum_m (-\omega_m - \frac{\Delta}{2})q_m(t) + g_m\sinh\theta_m\chi(t) \label{DymEqA}
\end{align}
applying the following transformation:
\begin{align}
& \chi(t) = \tilde{\chi}(t)\exp(-i\frac{\Delta}{2}t) \notag\\
&  p_m(t) = \tilde{p}_m(t)\exp\bigg[-i\left(\omega_m - \frac{\Delta}{2}\right)t\bigg] \notag\\
&  q_m(t) = \tilde{q}_m(t)\exp\bigg[-i\left(-\omega_m - \frac{\Delta}{2}\right)t\bigg]
\end{align}
we have:
\begin{align}
& i\frac{d\chi(t)}{dt} = \sum_m {g\cosh\theta_m}\tilde{p}_m(t)\exp[-i(\omega_m-\Delta)t] \notag\\
&+{g_m\sinh\theta_m}\tilde{q}_m(t)\exp[-i(-\omega_m-\Delta)t] \notag\\
& i\frac{d\tilde{p}(t)}{dt} = \sum_m {g_m\cosh\theta_m} \tilde{\chi}(t)\exp[i(\omega_m -\Delta)t] \notag\\
& i\frac{d\tilde{q}(t)}{dt} = \sum_m{g_m\sinh\theta_m} \tilde{\chi}(t)\exp[i(-\omega_m -\Delta)t] \label{DymEqB}
\end{align}

Integrating last two equations in (\ref{DymEqB}) and substituting into the first one, we have:
\begin{align}
 &\frac{d\tilde{\chi}(t)}{dt} = \notag\\
 &- \sum_m\cosh^2\theta_mg_m^2\int_0^t \tilde{\chi}(t')\exp[-i(\omega_m-\Delta)(t-t')]dt' \notag\\
   &- \sinh^2\theta_mg_m^2\int_0^t \tilde{\chi}(t')\exp[-i(-\omega_m-\Delta)(t-t')]dt'
\end{align}

The solution of the integro-differential equation above can be obtained iteratively. Since we are limited  to the frequent measurement limit where $g\tau < 1$,  only the short time behavior of $\tilde{\chi}(t)$ is concerned. This can be obtained by taking first iteration and inserting the initial condition  $\chi(0) =1$:
\begin{align}
 &\tilde{\chi}(t)\approx 1 - \sum_m\int_0^t (t-t') \cos^2\theta_mg_m^2\exp[-i(\omega -\Delta)t']dt'\notag\\
  &-  \sum_m\int_0^t (t-t') \sinh^2\theta_mg_m^2\exp[-i(-\omega -\Delta)t']dt' \notag\\
  & \approx  \exp\bigg\{ -t\sum_m g_m^2\big[{\cosh^2\theta_m}\mathcal{F}(\Delta,\omega) +{\sinh^2\theta_m}\mathcal{F}(\Delta,-\omega)\big]\bigg\} \label{ExpChiApp}
\end{align}
where
\begin{align}
  &\mathcal{F}(\Delta,\omega) = \frac{2\sin^2(\frac{\omega-\Delta}{2})t}{t(\omega-\Delta)^2} -i\frac{(\omega-\Delta)t-\sin(\omega-\Delta)t}{t(\omega-\Delta)^2}
\end{align}

Substituting (\ref{ExpChiApp}) into the definition of the decay rate of the survival probability (\ref{GammaLog}) , we finally obtain:

\begin{align}
  \gamma_{\text{th}}(\tau) =& \tau \sum_{m=0}^M\cosh^2 \theta_m g_m^2\text{sinc}^2\left[\frac{\tau(\omega_k-\Delta)}{2}\right] \notag\\
  +&\sinh^2\theta_m g_m^2\text{sinc}^2\left[\frac{\tau(\omega_k+\Delta)}{2}\right]
\end{align}
where $\text{sinc(x)}\equiv \sin(x)/x$.

\section{MPS-based numerically exact method }
As derived in the previous Appendix, the evolution of the MQRM from a squeezed thermal states can be described by the TFD schr\"{o}dinger equation of the Hamiltonian ($\ref{Ham3}$) with the initial state ($\ref{Inistate2}$). To implement MPS-based numerical method, the Hamiltonian is rewritten in a compact form of the following matrix product operator(MPO)\cite{wall2016simulating}:

\begin{align}
\hat{H} & =\left(\prod_{i=0}^{M-1}\mathcal{W}_b^{[i]}\right)\mathcal{W}_s \left(\prod_{j=0}^{M-1}\mathcal{W}_a^{[j]}\right) \notag\\
\mathcal{W}_b^{[m]} &=
  \left(
    \begin{array}{ccc}
      I &0  & 0 \\
      0 & I & 0 \\
      H_{b_m}, & G_{b_m}, & I \\
    \end{array}
  \right)\notag\\
  \mathcal{W}_s  &=
  \left(
    \begin{array}{ccc}
      I &0  & 0 \\
      \sigma_x & I & 0 \\
      \Delta\sigma_z/2, & \sigma_x, & I \\
    \end{array}
  \right)\notag\\
  \mathcal{W}_a^{[m]} &=
  \left(
    \begin{array}{ccc}
      I &0  & 0 \\
      G_{a_m} & I & 0 \\
      H_{a_m}, & 0, & I \\
    \end{array}
  \right)
\end{align}
where $I$ is the identity operator of the TLS and the boson space
\begin{align}
& H_{a_m} = A\omega_m a_m^\dag a_m + Ba_m^{\dag 2} + B^*a_m^{2} \notag\\
& H_{b_m} = -A\omega_m b_m^\dag b_m - Ba_m^{ 2} + B^*a_m^{\dag 2} \notag\\
& G_{a_m} = g_mKa_m^\dag\cosh\theta_m +\text{H.c} \notag \\
 &G_{b_m} = g_mKb\sinh\theta_m +\text{H.c}
\end{align}

\begin{figure}[tbp]
\includegraphics[scale=0.6]{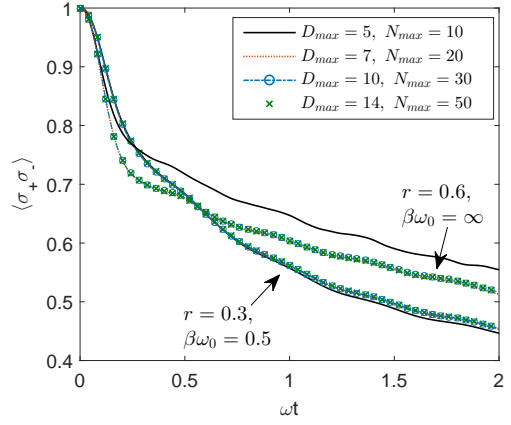}
\caption{(Color online)  The convergence of the MPS-TDVP method for the population evolution of the MQRM. Other parameters are $g/\omega_0 = 0.2, \phi = \pi/2, M =15$.}
\label{Fig5}
\end{figure}

Note for the open boundary condition, only first row of  $\mathcal{W}_b^{[1]}$ and first column of  $\mathcal{W}_a^{[M]}$ are used. With this MPOs,  we employ the recently proposed  time-dependent variational principle(TDVP) method\cite{haegeman2016unifying} to simulate the evolution. For numerical results throughout this study, the convergence can be achieved by setting the maximal bound dimensions of the MPS $D_{\text{max}} \le 15$ and the cutoff boson number for each cavity mode  $N_{\text{max}} \le 80$. The time step of simulations throughout this paper is set to $g\delta t = 10^{-3}$.

\newpage

\bibliographystyle{unsrt}%
\bibliography{MultiRabi_v2}

\begin{thebibliography}{10}

\bibitem{misra1977zeno}
Baidyanath Misra and EC~GEORGE Sudarshan.
\newblock The zeno¡¯s paradox in quantum theory.
\newblock {\em Journal of Mathematical Physics}, 18(4):756--763, 1977.

\bibitem{facchi2008quantum}
Paolo Facchi and Saverino Pascazio.
\newblock Quantum zeno dynamics: mathematical and physical aspects.
\newblock {\em Journal of Physics A: Mathematical and Theoretical},
  41(49):493001, 2008.

\bibitem{francica2010quantum}
F~Francica, F~Plastina, and S~Maniscalco.
\newblock Quantum zeno and anti-zeno effects on quantum and classical
  correlations.
\newblock {\em Physical Review A}, 82(5):052118, 2010.

\bibitem{schafer2014experimental}
Florian Sch{\"a}fer, Ivan Herrera, Shahid Cherukattil, Cosimo Lovecchio,
  Francesco~Saverio Cataliotti, Filippo Caruso, and Augusto Smerzi.
\newblock Experimental realization of quantum zeno dynamics.
\newblock {\em Nature communications}, 5, 2014.

\bibitem{itano1990quantum}
Wayne~M Itano, Daniel~J Heinzen, JJ~Bollinger, and DJ~Wineland.
\newblock Quantum zeno effect.
\newblock {\em Physical Review A}, 41(5):2295, 1990.

\bibitem{bernu2008freezing}
Julien Bernu, Samuel Del{\'e}glise, Cl{\'e}ment Sayrin, Stefan Kuhr, Igor
  Dotsenko, Michel Brune, Jean-Michel Raimond, and Serge Haroche.
\newblock Freezing coherent field growth in a cavity by the quantum zeno
  effect.
\newblock {\em Physical review letters}, 101(18):180402, 2008.

\bibitem{alvarez2010zeno}
Gonzalo~A {\'A}lvarez, DD~Bhaktavatsala Rao, Lucio Frydman, and Gershon
  Kurizki.
\newblock Zeno and anti-zeno polarization control of spin ensembles by induced
  dephasing.
\newblock {\em Physical review letters}, 105(16):160401, 2010.

\bibitem{pouyandeh2014measurement}
Sima Pouyandeh, Farhad Shahbazi, and Abolfazl Bayat.
\newblock Measurement-induced dynamics for spin-chain quantum communication and
  its application for optical lattices.
\newblock {\em Physical Review A}, 90(1):012337, 2014.

\bibitem{bayat2015measurement}
Abolfazl Bayat and Yasser Omar.
\newblock Measurement-assisted quantum communication in spin channels with
  dephasing.
\newblock {\em New Journal of Physics}, 17(10):103041, 2015.

\bibitem{barenco1997stabilization}
Adriano Barenco, Andre Berthiaume, David Deutsch, Artur Ekert, Richard Jozsa,
  and Chiara Macchiavello.
\newblock Stabilization of quantum computations by symmetrization.
\newblock {\em SIAM Journal on Computing}, 26(5):1541--1557, 1997.

\bibitem{beige2000quantum}
Almut Beige, Daniel Braun, Ben Tregenna, and Peter~L Knight.
\newblock Quantum computing using dissipation to remain in a decoherence-free
  subspace.
\newblock {\em Physical Review Letters}, 85(8):1762, 2000.

\bibitem{erez2008thermodynamical}
Noam Erez, Goren Gordon, Mathias Nest, and Gershon Kurizki.
\newblock Thermodynamical control by frequent quantum measurements.
\newblock {\em arXiv preprint arXiv:0804.2178}, 2008.

\bibitem{kakuyanagi2015observation}
K~Kakuyanagi, T~Baba, Y~Matsuzaki, H~Nakano, S~Saito, and K~Semba.
\newblock Observation of quantum zeno effect in a superconducting flux qubit.
\newblock {\em New Journal of Physics}, 17(6):063035, 2015.

\bibitem{forn2017ultrastrong}
P~Forn-D{\'\i}az, Juan~Jos{\'e} Garc{\'\i}a-Ripoll, Borja Peropadre, J-L
  Orgiazzi, MA~Yurtalan, R~Belyansky, Christopher~M Wilson, and A~Lupascu.
\newblock Ultrastrong coupling of a single artificial atom to an
  electromagnetic continuum in the nonperturbative regime.
\newblock {\em Nature Physics}, 13(1):39, 2017.

\bibitem{bosman2017multi}
Sal~J Bosman, Mario~F Gely, Vibhor Singh, Alessandro Bruno, Daniel Bothner, and
  Gary~A Steele.
\newblock Multi-mode ultra-strong coupling in circuit quantum electrodynamics.
\newblock {\em npj Quantum Information}, 3(1):46, 2017.

\bibitem{rabi1936process}
II~Rabi.
\newblock On the process of space quantization.
\newblock {\em Physical Review}, 49(4):324, 1936.

\bibitem{sundaresan2015beyond}
Neereja~M Sundaresan, Yanbing Liu, Darius Sadri, L{\'a}szl{\'o}~J Sz{\H{o}}cs,
  Devin~L Underwood, Moein Malekakhlagh, Hakan~E T{\"u}reci, and Andrew~A
  Houck.
\newblock Beyond strong coupling in a multimode cavity.
\newblock {\em Physical Review X}, 5(2):021035, 2015.

\bibitem{gely2017convergence}
Mario~F Gely, Adrian Parra-Rodriguez, Daniel Bothner, Ya~M Blanter, Sal~J
  Bosman, Enrique Solano, and Gary~A Steele.
\newblock Convergence of the multimode quantum rabi model of circuit quantum
  electrodynamics.
\newblock {\em Physical Review B}, 95(24):245115, 2017.

\bibitem{sanchez2018resolution}
Carlos S{\'a}nchez~Mu{\~n}oz, Franco Nori, and Simone De~Liberato.
\newblock Resolution of superluminal signalling in non-perturbative cavity
  quantum electrodynamics.
\newblock {\em Nature Communications}, 9, 2018.

\bibitem{de2014light}
Simone De~Liberato.
\newblock Light-matter decoupling in the deep strong coupling regime: The
  breakdown of the purcell effect.
\newblock {\em Physical review letters}, 112(1):016401, 2014.

\bibitem{garcia2015light}
Juan~Jose Garcia-Ripoll, Borja Peropadre, and Simone De~Liberato.
\newblock Light-matter decoupling and $a^2$ term detection in superconducting
  circuits.
\newblock {\em Scientific reports}, 5:16055, 2015.

\bibitem{lizuain2010zeno}
Ion Lizuain, Jorge Casanova, Juan~Jos{\'e} Garc{\'\i}a-Ripoll, JG~Muga, and
  Enrique Solano.
\newblock Zeno physics in ultrastrong-coupling circuit qed.
\newblock {\em Physical Review A}, 81(6):062131, 2010.

\bibitem{haegeman2016unifying}
Jutho Haegeman, Christian Lubich, Ivan Oseledets, Bart Vandereycken, and Frank
  Verstraete.
\newblock Unifying time evolution and optimization with matrix product states.
\newblock {\em Physical Review B}, 94(16):165116, 2016.

\bibitem{malekakhlagh2017cutoff}
Moein Malekakhlagh, Alexandru Petrescu, and Hakan~E T{\"u}reci.
\newblock Cutoff-free circuit quantum electrodynamics.
\newblock {\em Physical review letters}, 119(7):073601, 2017.

\bibitem{suzuki1991density}
Masuo Suzuki.
\newblock Density matrix formalism, double-space and thermo field dynamics in
  non-equilibrium dissipative systems.
\newblock {\em International Journal of Modern Physics B}, 5(11):1821--1842,
  1991.

\bibitem{gelin2017thermal}
Maxim~F Gelin and Raffaele Borrelli.
\newblock Thermal schr{\"o}dinger equation: Efficient tool for simulation of
  many-body quantum dynamics at finite temperature.
\newblock {\em Annalen der Physik}, 529(12):1700200, 2017.

\bibitem{schollwock2011density}
Ulrich Schollw{\"o}ck.
\newblock The density-matrix renormalization group in the age of matrix product
  states.
\newblock {\em Annals of Physics}, 326(1):96--192, 2011.

\bibitem{wall2016simulating}
Michael~L Wall, Arghavan Safavi-Naini, and Ana~Maria Rey.
\newblock Simulating generic spin-boson models with matrix product states.
\newblock {\em Physical Review A}, 94(5):053637, 2016.

\bibitem{schroder2016simulating}
Florian~AYN Schr{\"o}der and Alex~W Chin.
\newblock Simulating open quantum dynamics with time-dependent variational
  matrix product states: Towards microscopic correlation of environment
  dynamics and reduced system evolution.
\newblock {\em Physical Review B}, 93(7):075105, 2016.

\bibitem{kloss2018time}
Benedikt Kloss, Yevgeny~Bar Lev, and David Reichman.
\newblock Time-dependent variational principle in matrix-product state
  manifolds: Pitfalls and potential.
\newblock {\em Physical Review B}, 97(2):024307, 2018.

\bibitem{facchi2001quantum}
P~Facchi, H~Nakazato, and S~Pascazio.
\newblock From the quantum zeno to the inverse quantum zeno effect.
\newblock {\em Physical Review Letters}, 86(13):2699, 2001.

\bibitem{maniscalco2006zeno}
Sabrina Maniscalco, Jyrki Piilo, and Kalle-Antti Suominen.
\newblock Zeno and anti-zeno effects for quantum brownian motion.
\newblock {\em Physical review letters}, 97(13):130402, 2006.

\bibitem{he2017zeno}
Shu He, Qing-Hu Chen, and Hang Zheng.
\newblock Zeno and anti-zeno effect in an open quantum system in the
  ultrastrong-coupling regime.
\newblock {\em Physical Review A}, 95(6):062109, 2017.

\bibitem{zheng2008quantum}
H~Zheng, SY~Zhu, and MS~Zubairy.
\newblock Quantum zeno and anti-zeno effects: without the rotating-wave
  approximation.
\newblock {\em Physical review letters}, 101(20):200404, 2008.

\bibitem{cao2010dynamics}
Xiufeng Cao, JQ~You, H~Zheng, AG~Kofman, and Franco Nori.
\newblock Dynamics and quantum zeno effect for a qubit in either a low-or
  high-frequency bath beyond the rotating-wave approximation.
\newblock {\em Physical Review A}, 82(2):022119, 2010.

\bibitem{chaudhry2014zeno}
Adam~Zaman Chaudhry and Jiangbin Gong.
\newblock Zeno and anti-zeno effects on dephasing.
\newblock {\em Physical Review A}, 90(1):012101, 2014.

\bibitem{zhang2015zeno}
Yu-Ran Zhang and Heng Fan.
\newblock Zeno dynamics in quantum open systems.
\newblock {\em Scientific reports}, 5:11509, 2015.

\bibitem{wu2017quantum}
Wei Wu and Hai-Qing Lin.
\newblock Quantum zeno and anti-zeno effects in quantum dissipative systems.
\newblock {\em Physical Review A}, 95(4):042132, 2017.

\bibitem{schulman1994characteristic}
LS~Schulman, A~Ranfagni, and D~Mugnai.
\newblock Characteristic scales for dominated time evolution.
\newblock {\em Physica Scripta}, 49(5):536, 1994.

\bibitem{umezawa1982thermo}
Hiroomi Umezawa, Hiroshi Matsumoto, and Masashi Tachiki.
\newblock Thermo field dynamics and condensed states.
\newblock 1982.

\bibitem{borrelli2016quantum}
Raffaele Borrelli and Maxim~F Gelin.
\newblock Quantum electron-vibrational dynamics at finite temperature: Thermo
  field dynamics approach.
\newblock {\em The Journal of chemical physics}, 145(22):224101, 2016.

\bibitem{borrelli2017simulation}
Raffaele Borrelli and Maxim~F Gelin.
\newblock Simulation of quantum dynamics of excitonic systems at finite
  temperature: an efficient method based on thermo field dynamics.
\newblock {\em Scientific reports}, 7(1):9127, 2017.

\bibitem{chen2017finite}
Lipeng Chen and Yang Zhao.
\newblock Finite temperature dynamics of a holstein polaron: The thermo-field
  dynamics approach.
\newblock {\em The Journal of chemical physics}, 147(21):214102, 2017.

\bibitem{wang2017finite}
Lu~Wang, Yuta Fujihashi, Lipeng Chen, and Yang Zhao.
\newblock Finite-temperature time-dependent variation with multiple davydov
  states.
\newblock {\em The Journal of chemical physics}, 146(12):124127, 2017.

\bibitem{he2018zeno}
Shu He, Chen Wang, Li-Wei Duan, and Qing-Hu Chen.
\newblock Zeno effect of an open quantum system in the presence of 1/f noise.
\newblock {\em Physical Review A}, 97(2):022108, 2018.

\bibitem{houck2008controlling}
AA~Houck, JA~Schreier, BR~Johnson, JM~Chow, Jens Koch, JM~Gambetta,
  DI~Schuster, L~Frunzio, MH~Devoret, SM~Girvin, et~al.
\newblock Controlling the spontaneous emission of a superconducting transmon
  qubit.
\newblock {\em Physical review letters}, 101(8):080502, 2008.

\bibitem{mundarain2006zeno}
DF~Mundarain and J~Stephany.
\newblock Zeno and anti-zeno effect for a two-level system in a squeezed bath.
\newblock {\em Physical Review A}, 73(4):042113, 2006.

\bibitem{suzuki1986thermo}
Masuo Suzuki.
\newblock Thermo field dynamics of quantum spin systems.
\newblock {\em Journal of statistical physics}, 42(5-6):1047--1070, 1986.

\bibitem{suzuki1985thermo}
Masuo Suzuki.
\newblock Thermo field dynamics in equilibrium and non-equilibrium interacting
  quantum systems.
\newblock {\em Journal of the Physical Society of Japan}, 54(12):4483--4485,
  1985.

\bibitem{takahashi1996thermo}
Yasushi Takahashi and Hiroomi Umezawa.
\newblock Thermo field dynamics.
\newblock {\em International journal of modern Physics B},
  10(13n14):1755--1805, 1996.

\end{thebibliography}

\end{document}